\documentclass[prl,twocolumn,showpacs]{revtex4-1}
\usepackage{subfigure,color}
\usepackage{textcomp}
\usepackage{graphicx}
\usepackage{amssymb}
\usepackage{amsmath}
\usepackage{bm}
\usepackage{amsmath, amsthm, amssymb}

\begin{document}

\newcommand{\vk}{{\bf k}}
\def\ns{^{\vphantom{*}}}
\def\expect#1#2#3{{\langle #1 |   #2  |  #3 \rangle}}
\def\cH{{\cal H}}
\def\half{\frac{1}{2}}
\def\sut{\textsf{SU}(2)}
\def\suto{\textsf{SU}(2)\ns_1}
\def\kF{\ket{\,{\rm F}\,}}
\newcommand{\ack}[1]{[{\bf Pfft!: {#1}}]}
\newcommand{\bra}[1]{\langle #1 |}
\newcommand{\ket}[1]{| #1 \rangle}
\newcommand{\braket}[2]{\left \langle #1 | #2 \right\rangle}
\newcommand\R{{\mathrm {I\!R}}}
\newcommand\N{{\mathrm {I\!N}}}
\newcommand\h{{\cal H}}
\newcommand{\ra}{{\rightarrow}}
\newcommand{\be}{\begin{equation}}
\newcommand{\ee}{\end{equation}}
\newcommand{\bma}{\begin{pmatrix}}
\newcommand{\ema}{\end{pmatrix}}
\newcommand{\bali}{\begin{align}}
\newcommand{\eali}{\end{align}}

\newcommand{\ca}{\mathcal{A}}
\newcommand{\cf}{\mathcal{F}}
\newcommand{\bZ}{\mathbb{Z}}
\newcommand{\bI}{\mathbb{I}}
\newcommand{\ba}{\begin{eqnarray}}
\newcommand{\ea}{\end{eqnarray}}
\newcommand\tr{{\mbox{Tr\,}}}
\newcommand{\ignore}[1]{}
\newcommand{\pb}{{\textbf{p}}}
\newcommand{\pdag}{\phantom{\dagger}}

\title{Stabilization of Majorana modes in vortices in the superconducting phase of  topological insulators using topologically trivial bands.}


\author{Ching-Kai Chiu}
\affiliation{Department of Physics, University of Illinois, Urbana, IL 61801}

\author{Pouyan Ghaemi}
\affiliation{Department of Physics, University of Illinois, Urbana, IL 61801}

\author{Taylor L. Hughes}
\affiliation{Department of Physics, University of Illinois, Urbana, IL 61801}

\begin{abstract}
 It has been shown that doped topological insulators, up to certain level of doping, still preserve some topological signatures of the insulating phase such as axionic electromagnetic response and the the presence of a Majorana mode in the vortices of a superconducting phase.  Multiple topological insulators such as  HgTe, ScPtBi, and other ternary Heusler compounds  have been identified and generically feature the presence of a  topologically trivial band between the two topological bands. In this letter we show that the presence of such a trivial band can stabilize the topological signature over a much wider range of doping. Specifically, we calculate the structure of vortex modes in the superconducting phase of doped topological insulators a model that capture the features of HgTe and the ternary Heusler compounds. We show that due to the hybridization with the trivial band, Majorana modes are preserved over a large, extended doping range for p-doping. In addition to presenting a viable system where much less fine-tuning is required to observe the Majorana modes, our analysis opens a route to  study other topological features of doped compounds that cannot be modeled using the simple Bi$_2$Se$_3$ Dirac model.
\end{abstract}

\maketitle

Solid-state realizations of Majorana fermions, particles which are their own-anti-particles, are much sought after for their promise of new fundamental phenomena and associated quantum computing applications\cite{Nayak:2008fk}. By now the list of candidate systems to realize these particles has grown quite long including: non-Abelian fractional quantum Hall states\cite{ReadGreenP+ipFQHE00}, chiral superconductors\cite{dassarma}, topological insulators in proximity to superconductors\cite{FuKaneSCProximity}, axion strings\cite{sato1}, and low-dimensional spin-orbit coupled semiconductors coupled to superconductors\cite{DasSarmaMajoranaSCSCJunction,Gil2010}.  One of the most promising directions relies on the interplay between topological insulators (TIs)\cite{HasanKaneReview} and s-wave superconductivity\cite{FuKaneSCProximity,pavan}. Two such routes are: (i)the trapping of Majorana bound states (MBS) in vortices found in TI/s-wave superconductor proximity heterostructures\cite{FuKaneSCProximity} (ii)vortex bound states of doped TIs which exhibit an intrinsic s-wave superconductor phase\cite{pavan}.   The second mechanism has perhaps greater potential, as things currently stand, since it does not require the fine-tuning of TI material properties to make the system a bulk insulator \cite{HasanKaneReview}. Indeed, there is already a candidate material:  Cu doped Bi$_2$Se$_3$ which is a superconductor below 3.8 K\cite{SuperconductingCuxBi2Se3,CuxBi2Se3ARPES} although the precise nature of the superconducting state is still undetermined.

The initial proximity effect proposal of Fu and Kane illustrates that Majorana zero-modes will be localized where superconducting vortex lines in TI-superconductor heterostructures intersect the topological insulator surface states\cite{FuKaneSCProximity}. The bound states are confined to the vortex cores but they penetrate into the nominally gapped bulk-region below the surface. The localization length of the bound states in the bulk naturally depends on the size of the bulk gap and if the bulk is doped then the bound states can tunnel away from the surface and be destroyed\cite{pavan}. While this system, in most respects, is a trivial s-wave superconductor, it still remembers it  arose from a TI parent state as long as the doping is not too high. The consequence of this topological \emph{signature} is that vortices will still bind MBSs at the intersection between the vortex line and the surface. As the TI becomes more doped  a vortex quantum phase transition (VQPT) occurs at a critical chemical potential and beyond which no Majorana modes are bound to the vortices. 
This mechanism comes with its own challenges as one desires the formation of an s-wave gap (which may or may not be favored), and there is only a finite doping range over which the system will exhibit this phenomenon. The latter restriction leads to the competition between a desire to stabilize a superconducting phase (higher doping) and preserving the MBSs (lower-doping) and thus  might still require fine-tuning of the chemical potential. 

Here we find a more attractive material context for this mechanism which removes the fine-tuning constraint by considering materials with electronic structure similar to bulk HgTe. This requires the consideration of 6-bands instead of the 4-band model of the Bi$_2$Se$_3$/Bi$_2$Te$_3$ family\cite{SCZhangBi2X3DiracCone}.  Our essential insight is that the coupling of the Dirac structure to an additional ``trivial" band(s) can strongly affect the location of the critical chemical potential and stabilize the MBSs over a \emph{much larger} range of doping by extending the energy range in which the topological signature persists. The set of materials where our analysis can be applied is quite large and includes bulk HgTe, which was recently confirmed to be a TI\cite{HgTeTI}, and the ternary Heusler materials (\emph{e.g.} ScPtBi) which were recently predicted to be TIs\cite{Heusler}.  Beyond this application our results indicate that the presence of a topologically trivial band can aid the  persistence of topological features (\emph{e.g.} magneto-electric response) beyond the un-doped regime. For these materials this implies the existence of a unique metallic phase over a broad doping range that has not been fully characterized.


In this letter we will primarily consider bulk HgTe, a zero-gap semiconductor that exhibits a TI phase under compressive strain although our results directly apply to the ternary Heusler compounds with just a change of parameters in the model\cite{FuKaneTIInversion,dai2008,HgTeTI}.  The low-energy band structure is time-reversal invariant and consists of three doubly-degenerate bands near the $\Gamma$-point. The bands closest to the Fermi-level nominally consist of the $J=3/2$ multiplet of a spin-orbit split p-orbital (denoted by $\Gamma_8$) which, when (un)strained is (4)2-fold degenerate at the $\Gamma$-point. The other relevant band nominally consists of an s-orbital ($\Gamma_6$) and forms a Dirac Hamiltonian with the $J_z=\pm 1/2$ states (the light hole(LH) band) of the $\Gamma_8$ multiplet\cite{dai2008}.  The relative orientation of the $\Gamma_6$ and $\Gamma_8$ bands are inverted and thus induce a topological band structure \cite{FuKaneTIInversion}. The other doubly degenerate band, which is made from the $J_z=\pm 3/2$ states (the heavy hole (HH) band) hybridizes with both of the Dirac bands, but does not qualitatively affect the topological properties\cite{dai2008}.   The 6-band model Hamiltonian expanded near the $\Gamma$-point in a basis convenient for our analysis  $\left[\ket{\Gamma_8,3/2},\ \ket{\Gamma_6,1/2},\ \ket{\Gamma_8,-1/2},\right.$ $\ket{\Gamma_8,-3/2},$ $\left. \ket{\Gamma_6,-1/2},\ \ket{\Gamma_8,1/2}\right]$ is:
\begin{align}
H_{\mathrm{HgTe}}(\vec{k})=
\begin{pmatrix}
H_+(\vec{k}) & H_c(\vec{k}) \\
H^{\dagger}_c(\vec{k})  & H_-(\vec{k})
\end{pmatrix},\label{eq:Ham}
\end{align}
where
\begin{footnotesize}
\begin{align}
&H_+(\vec{k})= \nonumber \\
&\begin{pmatrix}
 E_\nu-A_1k_{\parallel}^2-A_2k_z^2-\Sigma& -\lambda\frac{P}{\sqrt{2}}k_- &  \sqrt{3}\lambda (Dk_-^2-F k_+^2)  \nonumber \\
-\lambda\frac{P}{\sqrt{2}}k_+ & A_0 k^2  & \frac{P}{\sqrt{6}} k_-  \\
\sqrt{3}\lambda(Dk_+^2-Fk_-^2) & \frac{P}{\sqrt{6}}k_+ & E_\nu-B_1k_{\parallel}^2-B_2k_z^2+\Sigma
\end{pmatrix} \nonumber \\
&H_c(\vec{k})= 
\begin{pmatrix}
0 & 0 & 2\sqrt{3}Gk_-k_z \\
0 & 0 & \sqrt{\frac{2}{3}}Pk_z  \\
 -2\sqrt{3}Gk_-k_z  &  \sqrt{\frac{2}{3}}Pk_z & 0
\end{pmatrix} ,\ \\
&H_-(\vec{k})=H_+^*(-\vec{k}),\ k_{\parallel}^2=k_x^2+k_y^2,\ k_\pm=k_x\pm ik_y. \nonumber
\end{align}
\end{footnotesize}
\noindent where we have introduced a model parameter $\lambda$ which we will use to tune the coupling between the HH bands and the other bands to pass from the Dirac limit ($\lambda=0$) to the physical limit ($\lambda=1$). Additionally, $\Sigma$ is the gap generated by the compressive strain, and the remaining parameters are determined by the material band structure\cite{Novik:2005fk}. For all of the materials in which we are interested $E_\nu>0$ and $\Sigma>0.$ The Hamiltonian $H_{{\mathrm{HgTe}}}$ preserves time reversal symmetry with the operator $T=-i\sigma_y\otimes \bI_{3\times 3}\Theta$ where $\Theta$ is complex conjugation. When $k_z=0$, $H_{{\mathrm{HgTe}}}$ is block diagonal with two decoupled $3\times 3$ blocks which are time-reversed partners. The energy spectra and orbital composition when $\lambda = 0 (1)$ are presented in Fig. 1a(d). 

\begin{figure*}[t]\
\includegraphics[width=16cm]{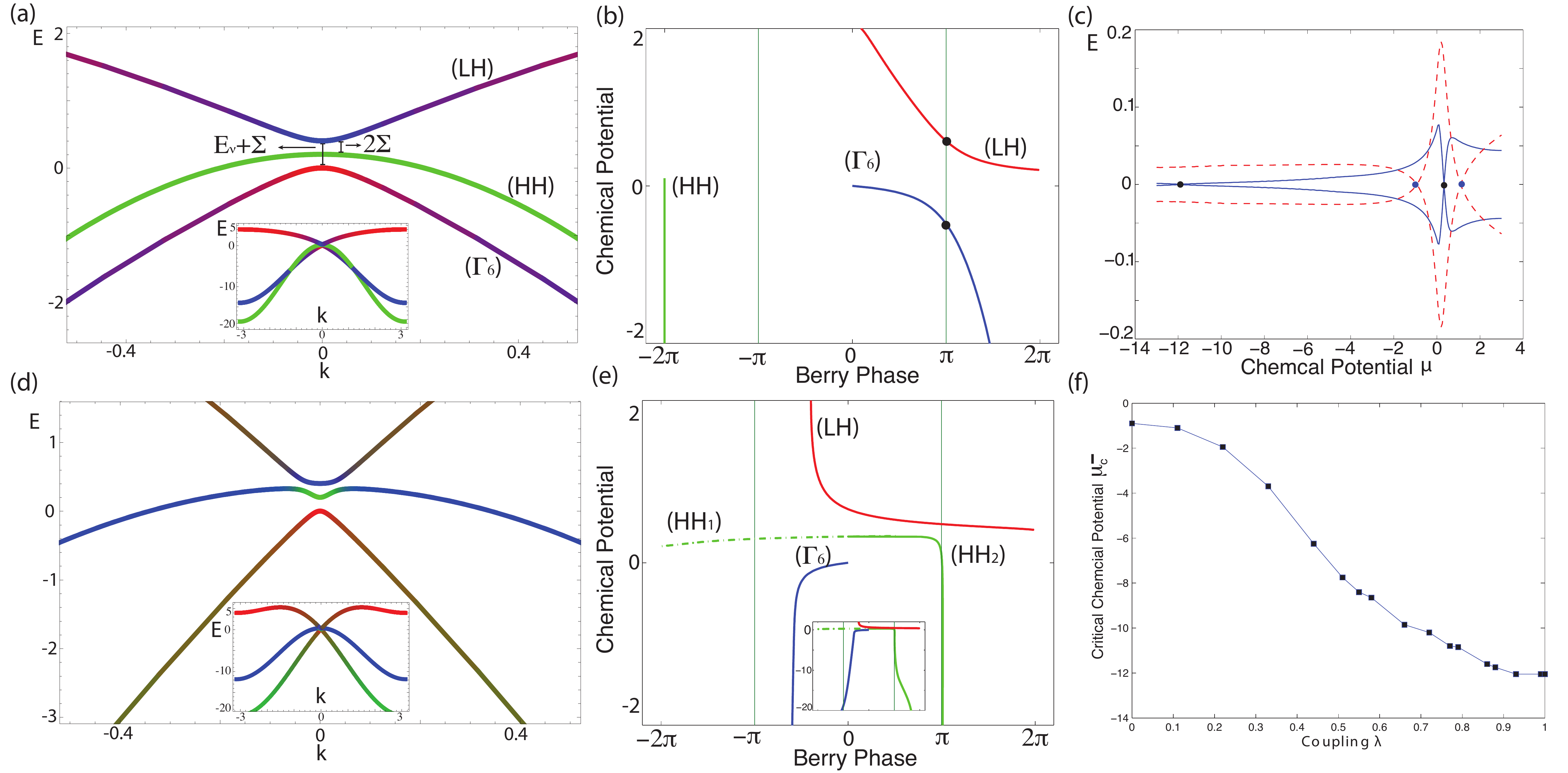}
\caption{Band structure and orbital content of the bands of HgTe (indicated by colors: {blue $\ket{\Gamma_8,\pm 1/2}$, red $\ket{\Gamma_6,\pm 1/2}$, green $\ket{\Gamma_8,\pm 3/2}$}) for $\lambda=0$ (a) and $\lambda=1$ (d). Insets show zoomed out energy spectrum. Corresponding Berry-phase on the Fermi surfaces at $k_z=0$ as a function of chemical potential for $\lambda=0$ (b) and $\lambda=1$ (e). The Berry phases are calculated by a numerical line-integral around each independent Fermi-surface. The dashed green/solid green lines in (e) represent the two different HH Fermi-surfaces. Inset shows zoomed out Berry phase so that the $\pi$ Berry phase of the $\Gamma_6$ band can be seen. (c) Exact diagaonlization results for the lowest energy vortex core states for $\lambda=0,1$ dashed/solid respectively where dots indicate critical points. This was calculated for $k_z=0$ for a lattice size of $180\times 180$ with HgTe parameters adapted from Ref. \onlinecite{Novik:2005fk} and $\Sigma=0.1, \Delta=0.3.$ (f) Valence-band $\mu_{c}^{-}$ as a function of $\lambda.$}
\end{figure*}

To determine the existence of MBSs on superconductor vortices we will add s-wave superconductivity to the 6-band model at the mean-field level and study the Bougoliobov-De Gennes (BdG) spectrum when a vortex line is oriented along the z-direction. Note that we are not considering proximity induced superconductivity but s-wave superconductivity generated due to a Fermi-surface instability in doped topological insulators\cite{pavan}. Besides the BdG doubling of the Hamiltonian, one needs to include the vortex winding in the phase of the superconducting order parameter. Considering the extreme type-II superconductor limit we ignore the magnetic field which does not qualitatively change the results \cite{Chiu:2011uq}. 
 In an ordinary s-wave superconductor, the vortex core states are gapped \cite{CdGMSCVortexStates64} with a ``mini"-gap $\delta\propto \frac{\Delta^2}{E_f}$ where $\Delta$ is the superconducting gap and $E_F$ is the Fermi energy of the normal metal state. Unexpectedly, it was found that in doped topological insulators, these modes can become gapless at a critical doping level $\mu_c$ leading to the aforementioned VQPT \cite{pavan}. Essentially, when tuned away from such a critical point, one can think of the vortex line itself as a gapped 1D superconductor which can either exist in a topological $(\mu_c^{-}<\mu<\mu^{+}_c)$ or trivial ($\mu>\mu^{+}_c$ or $\mu<\mu^{-}_c$ phase where  $ \mu^{\pm}_c$ is the critical chemical potential in the conduction and valance bands respectively) phase analogous to the Kitaev p-wave superconducting wire\cite{KitaevWireMajorana01,pavan}. 
When the vortex is in a topological phase it will have MBSs localized on its ends \emph{i.e.} at places where the vortex line terminates on a surface. This is exactly the phenomenon we seek and thus our goal is to find the region of chemical potential where the 1D vortex line is in a gapped topological phase. 
Here, instead of studying the appearance of Majorana modes on the boundary, we study the VQPT in a torus geometry with a vortex line  along the periodic z-direction and use Lanczos exact diagonalization calculations find the lowest energy vortex mode \cite{pavan,Chiu:2011uq}.

For comparison we show two different results, each for a  different coupling $\lambda=0, 1.$  In the case $\lambda=0$ (Fig.1c, dashed line) we essentially reproduce the results of Ref. \onlinecite{pavan} as this is the Dirac limit. We see two critical points one in the conduction band and the other in the valence band(s); both critical points occur past the onset of bulk doping. For $\lambda=1$ (Fig. 1c solid line) the HH band is fully coupled and we can see a critical point in the conduction band that has been pushed toward the conduction band edge while the critical point in the valence band has been driven to a \emph{much} lower energy \emph{i.e.} vastly increasing the range of valence band doping which would contain stable MBSs. 

This large effect is rather surprising, but we will now argue that it is the natural outcome of coupling to a trivial band.  From the arguments of Ref. \onlinecite{pavan} we know that in the \emph{weak-pairing} limit $(\Delta << E_F)$ we can determine the VQPT from the normal-state Hamiltonian alone, without using the BdG formalism. In this case, for a vortex-line oriented in the z-direction, the critical chemical potential is determined solely by the Berry phases of the normal-state Fermi-surface (FS) in the $k_z=0$ and $k_z=\pi$ planes. For models of the type we are considering there are no low-energy states near $k_z=\pi$ so we will only consider the $k_z=0$ plane. It was shown that for a single (spin degenerate) FS that { the lowest vortex core-state energies are proportional to $E_n\propto [\Phi_B +(2n-1)\pi]$, where n is an integer}, so the vortex-line is critical when the Berry-phase $\Phi_B$ around the FS is $\pi.$ Additionally, for a four-band Dirac Hamiltonian, there is an easy way to determine the energies for FS's with $\pi$ Berry phase as the $k$-dependent mass term vanishes exactly at that chemical potential. That is, at this chemical potential there is a VQPT point where a zero energy vortex mode emerges. For our $6$-band model there are some chemical potentials where there are more than one independent FS when doped into the valence band(s), but let us first consider cases where there is only a single (spin degenerate) FS.  While the the presence of zero energy mode directly depends on the the value of $\Phi_B$ \cite{pavan} even the single-FS case does not allow us to determine $\Phi_B=\pi$ in a simple way because the vanishing of the mass term of the Dirac block does not guarantee $\Phi_B=\pi$ when the HH band is added ($\lambda\neq 0$). To determine $\Phi_B$ we need to consider $k_z=0$ where Eq. \ref{eq:Ham} breaks up into two decoupled $3\times 3$ blocks. Let us consider one block of the Hamiltonian in the $k_x k_y$-plane in polar coordinates $(k_{\parallel},\theta)$ in the basis of $\left[\ket{m_J=3/2},\ \ket{m_J=1/2},\ \ket{m_J=-1/2}\right],$ 
\begin{small}
\begin{align}
\label{H1}
&H_+(k_{\parallel},\theta)= \nonumber \\
&\begin{pmatrix}
 E_\nu-A_1k_{\parallel}^2-\Sigma& -\frac{\lambda P}{\sqrt{2}}k_{\parallel}e^{-i\theta} &  \sqrt{3}D\lambda k_{\parallel}^2e^{-2i\theta} \\
-\frac{\lambda P}{\sqrt{2}}k_{\parallel}e^{i\theta} & A_0 k_{\parallel}^2  & \frac{P}{\sqrt{6}} k_{\parallel}e^{-i\theta}  \\
 \sqrt{3}D\lambda k_{\parallel}^2e^{2i\theta} & \frac{P}{\sqrt{6}}k_{\parallel}e^{i\theta} & E_\nu-B_1k_{\parallel}^2+\Sigma
\end{pmatrix}.
\end{align}
\end{small}\noindent 
Due to time reversal symmetry, the Berry phases of $H_-$ and $H_+$ only differ by signs so we only consider $H_+$ and calculate the Berry phases at the FS's.  To simplify the calculation, we have used the spherical approximation which implies that $A_1=A_2,\; B_1=B_2,$ and $F=0$ in Eq. \ref{eq:Ham}, and that in-plane FS's will be circular. The generic solution can be obtained by using an ansatz which captures the $\theta$-dependence and is of the form:
$\Psi=\left(a(k_{\parallel})e^{-i\theta},\; b(k_{\parallel}), \; c(k_{\parallel})e^{i\theta} \right)^{T}$ where the vector $(a,\ b,\ c)$ is an eigenvector of  $H_{+}(k_{\parallel},0).$
Therefore, for the case of a single in-plane FS, the Berry phase of the wavefunction $\Psi$ is 
\be
\Phi_B=-i\int_0^{2\pi}\bra{\Psi}\partial_\theta\ket{\Psi}d\theta=2\pi(|c|^2-|a|^2).
\ee This result applies in the entire regime of n-doping, and in several regions of p-doping where only a single HH FS persists. The expression for $\Phi_B$ is generic for our model and the weights $a,b,c$ depend on the particular FS, the value of $\mu,$ and the value of $\lambda.$

Let us now illustrate how turning-on the coupling to the the trivial HH band affects the locations of critical FS's \emph{i.e.} places where $\Phi_B=\pi.$ First, we consider $\lambda=0$ so that the system is in the Dirac limit (see Fig. 1b).  If the Fermi-level is in the HH band the Berry phase is an integer multiple of $2\pi$  (since $\vert a\vert^2=1, \vert c\vert^2=0$) and thus does not generate a VQPT. However, if the Fermi-level is in one of the two Dirac bands then the Berry phase is $\Phi_B=2\pi|c|^2$ which will reach a value of $\pi$ when $|c|^2=|b|^2=1/2$ as found in Ref. \cite{pavan}. This will occur at one of the chemical potentials $\mu^{\pm}_c(\lambda=0)=\frac{A_0(E_\nu+\Sigma)}{A_0+B_1}\pm\sqrt{\frac{P^2(E_\nu+\Sigma)}{6(A_0+B_1)}}$ for the upper and lower bands respectively as indicated by the solid dots in Fig. 1b. With a small $\lambda\neq 0$ there will be a correction to the LH and $\Gamma_6$ band wavefunctions of the form $\delta\Psi_{\pm}=(\delta_{\pm},0,0)^T$ where the $\pm$ indicate upper(LH)/lower ($\Gamma_6$) band respectively. The resulting wavefunctions and  Berry phases at the original critical energies are
$\Psi_\pm =(\delta_\pm,\ \pm 1,\ 1)^T/ \sqrt{2+|\delta_\pm|^2}$ and  
$\Phi_{B\pm} =\pi(1-3|\delta_\pm|^2/(2+|\delta_{\pm}|^2))$
which have clearly been decreased independent of which band. The effect on the respective critical potentials in each band is, however, asymmetric. 
We can see this by noting for $\lambda=0,$ $\Phi_B$ shifts from $2\pi$ to $0$ ($0$ to $2\pi$) as one raises (lowers) the chemical potential through the LH ($\Gamma_6$) band (see Fig. 1b).  Our restriction to a single-FS implies that the critical points are determined by $\Phi_B=\pi.$  Thus, with respect to the original critical points we see that the perturbed critical points are at $\mu^{\pm\ast}_{c}<\mu^{\pm}_{c}.$ This means that the viable doping range has \emph{decreased} for the conduction band, but \emph{increased} for the valence band. Indeed, we see it is exactly the coupling to the trivial band which causes the shift in the Berry phase. If we assume axial rotation symmetry one can easily extend this result to trivial bands with higher angular momentum by replacing $\ket{m_J=3/2}$ with $\ket{m_J=(n+1/2)}$ which will lead to $\Phi_B=2\pi(|c|^2-n|a|^2).$ { For $n\geq 0$}, this will share the same qualitative features as the $m_J=3/2$ case; {but for  $n<0$, the range of the MBSs will  \emph{decrease} for the \emph{valence} band and \emph{increase} for the \emph{conduction} band. } 
For most ranges of p-doping, when $\lambda\neq 0,$ this calculation is merely suggestive since there will be more than one FS, but we will see in the next paragraph that these arguments are still approximately valid in most circumstances. We also note that in Fig. 1c,f we have already seen that the critical point can be shifted dramatically in the valence band due to the nature of the coupling to the trivial HH band, but one must eventually take other bands into account, because as $\mu^{-}_{c}$ is pushed lower and lower we will eventually have to consider the spin-orbit split off band which we have ignored. The qualitative trend, however, is quite striking and is favorable for experimental realization.

Finally, to complete our analytic interpretation of the numerics we consider the important case when two FS's are present simultaneously for some range of chemical potentials. For our model there are two FS's present in two separate energy ranges (when $\lambda\neq 0$): (i) a chemical potential near the top of the HH band leading to a narrow energy range with two HH FS's (ii) a wide range of chemical potentials with one FS from the HH band and one from the $\Gamma_6$ band.  If momentum is approximately conserved in the presence of the vortex \emph{i.e.} if the vortex profile is smooth and broad, 
 we can treat the two FS approximately independently. For case (i) we can see from Fig. 1e that there are a total of two places where $\Phi_B=\pm \pi$ one for each FS. 
These critical points are quite unstable and usually couple together to annihilate. 
On the other hand, for case (ii) we see from Fig. 1e that in this energy range the HH FS will never have a point where $\Phi_B=\pi$ so the core states generated by this FS will always remain gapped. So in this regime, in the decoupled FS limit,  the VQPT will continue to be controlled by the FS of the $\Gamma_6$ band and we can continue to apply our previous analytic arguments above. 
To more accurately match our numerics we must consider the hybridization between the vortex core modes on the two FS's through perturbation theory. If we let $E_1$ ($E_2$) be the lowest energy core state from the HH ($\Gamma_6$) FS we can write down the effective Hamiltonian
\be
H_{\mathrm{eff}}=
\begin{pmatrix}
E_1 & \alpha^* & 0 & -\beta \\
\alpha & E_2 & \beta & 0 \\
0 & \beta^* & -E_1 & -\alpha \\
-\beta^* & 0 & -\alpha^* & -E_2
\end{pmatrix}, \label{Heff}
\ee
where $\alpha$ and $\beta$ are the only couplings allowed by the BdG particle-hole symmetry.  { For the decoupled limit $\alpha=\beta=0$, the lowest core-state energy from each single FS can be determined by the the Berry phase of the normal-state FS in the $k_z=0$ plane; for case(ii) $E_1\propto(\Phi_B^{\text{HH}}+\pi)> 0$ and $E_2\propto (\Phi_B^{\Gamma_6}+\pi)$, where $\Phi_B^{\text{HH}}$ and $\Phi_B^{\Gamma_6}$ are the Berry phases of the HH and $\Gamma_6$ FS's respectively. Therefore, $E_2=0$ exactly when $\mu=\mu^{-}_c$ so that $\Phi_B^{\Gamma_6}=-\pi$.} When the inter-FS couplings are on, $H_\mathrm{eff}$ has a pair of zero eigenvalues, \emph{i.e.} a critical point, when $E_1E_2=|\alpha|^2-|\beta|^2$. The values of $\alpha, \beta$ depend on the details of the vortex core states and are sensitive to finite-size effects. Since { $E_2$ decreases from positive to negative} as $\mu$ gets lower then if $|\alpha|>|\beta |$ ($|\alpha |< |\beta |$) the critical chemical potential is modified and driven toward (away from) the band edge. In our numerical simulations we find the former case where the critical chemical potential is reached \emph{before} a Berry phase $\Phi_B^{\Gamma_6}=-\pi$ is reached. While this effect moves the critical point back toward the band edge it still does not counter-act the much larger shift due to the hybridization between the $\Gamma_6$ and HH bands as we have seen in  Fig. 1c,f. We note in passing that we have also considered the effects due to bulk inversion asymmetry and find they do not qualitatively alter our results. 

Our calculations indicate that materials with Dirac bandstructures that hybridize with trivial bands can support vortex MBSs over much larger ranges of doping. The hybridization with the trivial band delays the VQPT in the valence band while accelerating it in the conduction band. We are optimistic that these effects can be used to find and design an ideal material to support Majorana vortex states. More generally we showed that the trivial band coupling extends the range of the topological signature. This merits further investigations on other features of doped TIs which are sensitive to the band inversion of the parent insulator state.

CKC is supported by the NSF under grant DMR 09-03291 and thanks  the Topo-Mat-11 program at KITP for its hospitality. PG is supported by the ICMT at UIUC. TLH is supported by U.S. DOE, Office of Basic
Energy Sciences, Division of Materials Sciences and Engineering under Award DE-FG02-07ER46453.


\begin{thebibliography}{10}%
\makeatletter
\providecommand \@ifxundefined [1]{%
 \ifx #1\undefined \expandafter \@firstoftwo
 \else \expandafter \@secondoftwo
\fi
}%
\providecommand \@ifnum [1]{%
 \ifnum #1\expandafter \@firstoftwo
 \else \expandafter \@secondoftwo
\fi
}%
\providecommand \enquote [1]{``#1''}%
\providecommand \bibnamefont  [1]{#1}%
\providecommand \bibfnamefont [1]{#1}%
\providecommand \citenamefont [1]{#1}%
\providecommand\href[0]{\@sanitize\@href}%
\providecommand\@href[1]{\endgroup\@@startlink{#1}\endgroup\@@href}%
\providecommand\@@href[1]{#1\@@endlink}%
\providecommand \@sanitize [0]{\begingroup\catcode`\&12\catcode`\#12\relax}%
\@ifxundefined \pdfoutput {\@firstoftwo}{%
 \@ifnum{\z@=\pdfoutput}{\@firstoftwo}{\@secondoftwo}%
}{%
 \providecommand\@@startlink[1]{\leavevmode}%
 \providecommand\@@endlink[0]{}%
}{%
 \providecommand\@@startlink[1]{%
  \leavevmode
  \pdfstartlink
   attr{/Border[0 0 1 ]/H/I/C[0 1 1]}%
   user{/Subtype/Link/A<</Type/Action/S/URI/URI(#1)>>}%
  \relax
 }%
 \providecommand\@@endlink[0]{\pdfendlink}%
}%
\providecommand \url  [0]{\begingroup\@sanitize \@url }%
\providecommand \@url [1]{\endgroup\@href {#1}{\urlprefix}}%
\providecommand \urlprefix [0]{URL }%
\providecommand \Eprint[0]{\href }%
\@ifxundefined \urlstyle {%
  \providecommand \doi [1]{doi:\discretionary{}{}{}#1}%
}{%
  \providecommand \doi [0]{doi:\discretionary{}{}{}\begingroup
  \urlstyle{rm}\Url }%
}%
\providecommand \doibase [0]{http://dx.doi.org/}%
\providecommand \Doi[1]{\href{\doibase#1}}%
\providecommand \bibAnnote [3]{%
  \BibitemShut{#1}%
  \begin{quotation}\noindent
    \textsc{Key:}\ #2\\\textsc{Annotation:}\ #3%
  \end{quotation}%
}%
\providecommand \bibAnnoteFile [2]{%
  \IfFileExists{#2}{\bibAnnote {#1} {#2} {\input{#2}}}{}%
}%
\providecommand \typeout [0]{\immediate \write \m@ne }%
\providecommand \selectlanguage [0]{\@gobble}%
\providecommand \bibinfo [0]{\@secondoftwo}%
\providecommand \bibfield [0]{\@secondoftwo}%
\providecommand \translation [1]{[#1]}%
\providecommand \BibitemOpen[0]{}%
\providecommand \bibitemStop [0]{}%
\providecommand \bibitemNoStop [0]{.\EOS\space}%
\providecommand \EOS [0]{\spacefactor3000\relax}%
\providecommand \BibitemShut [1]{\csname bibitem#1\endcsname}%
\bibitem{Nayak:2008fk}%
  \BibitemOpen
  \bibfield{author}{%
  \bibinfo {author} {\bibfnamefont{C.}~\bibnamefont{Nayak}}, \bibinfo {author}
  {\bibfnamefont{S.~H.}\ \bibnamefont{Simon}}, \bibinfo {author}
  {\bibfnamefont{A.}~\bibnamefont{Stern}}, \bibinfo {author}
  {\bibfnamefont{M.}~\bibnamefont{Freedman}},\ and\ \bibinfo {author}
  {\bibfnamefont{S.}~\bibnamefont{Das~Sarma}},\ }%
  \bibfield{journal}{%
  \bibinfo {journal} {Reviews of Modern Physics}\ }%
  \textbf{\bibinfo {volume} {80}},\ \bibinfo {pages} {1083} (\bibinfo {month}
  {09}\ \bibinfo {year} {2008})
  \bibAnnoteFile{NoStop}{Nayak:2008fk}%
\bibitem{ReadGreenP+ipFQHE00}%
  \BibitemOpen
  \bibfield{author}{%
  \bibinfo {author} {\bibfnamefont{N.}~\bibnamefont{Read}}\ and\ \bibinfo
  {author} {\bibfnamefont{D.}~\bibnamefont{Green}},\ }%
  \bibfield{journal}{%
  \Doi{10.1103/PhysRevB.61.10267}{\bibinfo {journal} {Phys. Rev. B}}\ }%
  \textbf{\bibinfo {volume} {61}},\ \bibinfo {pages} {10267} (\bibinfo {year}
  {2000})%
  \bibAnnoteFile{NoStop}{ReadGreenP+ipFQHE00}%
\bibitem{dassarma}%
  \BibitemOpen
  \bibfield{author}{%
  \bibinfo {author} {\bibfnamefont{S.}~\bibnamefont{Das~Sarma}}, \bibinfo
  {author} {\bibfnamefont{C.}~\bibnamefont{Nayak}},\ and\ \bibinfo {author}
  {\bibfnamefont{S.}~\bibnamefont{Tewari}},\ }%
  \bibfield{journal}{%
  \Doi{10.1103/PhysRevB.73.220502}{\bibinfo {journal} {Phys. Rev. B}}\ }%
  \textbf{\bibinfo {volume} {73}},\ \bibinfo {pages} {220502} (\bibinfo {year}
  {2006})%
  \bibAnnoteFile{NoStop}{dassarma}%
\bibitem{FuKaneSCProximity}%
  \BibitemOpen
  \bibfield{author}{%
  \bibinfo {author} {\bibfnamefont{L.}~\bibnamefont{Fu}}\ and\ \bibinfo
  {author} {\bibfnamefont{C.~L.}\ \bibnamefont{Kane}},\ }%
  \bibfield{journal}{%
  \Doi{10.1103/PhysRevLett.100.096407}{\bibinfo {journal} {Phys. Rev. Lett.}}\
  }%
  \textbf{\bibinfo {volume} {100}},\ \bibinfo {pages} {096407} (\bibinfo {year}
  {2008})%
  \bibAnnoteFile{NoStop}{FuKaneSCProximity}%
    \bibitem{sato1}%
  \BibitemOpen
  \bibfield{author}{%
  \bibinfo {author} {\bibfnamefont{M.}\ \bibnamefont{Sato}},\ }%
  \bibfield{journal}{%
  \Doi{10.1070/1063-7869/44/10S/S29}{\bibinfo {journal} {Physics Letters B}}\
  }%
  \textbf{\bibinfo {volume} {575}},\ \bibinfo {pages} {126} (\bibinfo {year}
  {2003})%
  \bibAnnoteFile{NoStop}{sato1}%
\bibitem{DasSarmaMajoranaSCSCJunction}%
  \BibitemOpen
  \bibfield{author}{%
  \bibinfo {author} {\bibfnamefont{R.~M.}\ \bibnamefont{Lutchyn}}, \bibinfo
  {author} {\bibfnamefont{J.~D.}\ \bibnamefont{Sau}},\ and\ \bibinfo {author}
  {\bibfnamefont{S.}~\bibnamefont{Das~Sarma}},\ }%
  \bibfield{journal}{%
  \Doi{10.1103/PhysRevLett.105.077001}{\bibinfo {journal} {Phys. Rev. Lett.}}\
  }%
  \textbf{\bibinfo {volume} {105}},\ \bibinfo {pages} {077001} (\bibinfo {year}
  {2010})%
  \bibAnnoteFile{NoStop}{DasSarmaMajoranaSCSCJunction}%
\bibitem{Gil2010}%
  \BibitemOpen
  \bibfield{author}{%
  \bibinfo {author} {\bibfnamefont{Y.}~\bibnamefont{Oreg}}, \bibinfo {author}
  {\bibfnamefont{G.}~\bibnamefont{Refael}},\ and\ \bibinfo {author}
  {\bibfnamefont{F.}~\bibnamefont{von Oppen}},\ }%
  \bibfield{journal}{%
  \Doi{10.1103/PhysRevLett.105.177002}{\bibinfo {journal} {Phys. Rev. Lett.}}\
  }%
  \textbf{\bibinfo {volume} {105}},\ \bibinfo {pages} {177002} (\bibinfo {year}
  {2010})%
  \bibAnnoteFile{NoStop}{Gil2010}%
\bibitem{HasanKaneReview}%
  \BibitemOpen
  \bibfield{author}{%
  \bibinfo {author} {\bibfnamefont{M.~Z.}\ \bibnamefont{Hasan}}\ and\ \bibinfo
  {author} {\bibfnamefont{C.~L.}\ \bibnamefont{Kane}},\ }%
  \bibfield{journal}{%
  \Doi{10.1103/RevModPhys.82.3045}{\bibinfo {journal} {Rev. Mod. Phys.}}\ }%
  \textbf{\bibinfo {volume} {82}},\ \bibinfo {pages} {3045} (\bibinfo {year}
  {2010})%
  \bibAnnoteFile{NoStop}{HasanKaneReview}%
\bibitem{pavan}%
  \BibitemOpen
  \bibfield{author}{%
  \bibinfo {author} {\bibfnamefont{P.}~\bibnamefont{Hosur}}, \bibinfo {author}
  {\bibfnamefont{P.}~\bibnamefont{Ghaemi}}, \bibinfo {author}
  {\bibfnamefont{R. S. K.}~\bibnamefont{Mong}},\ and\ \bibinfo {author}
  {\bibfnamefont{A.}~\bibnamefont{Vishwanath}},\ }%
  \bibfield{journal}{%
  \Doi{10.1103/PhysRevLett.107.097001}{\bibinfo {journal} {Phys. Rev. Lett.}}\
  }%
  \textbf{\bibinfo {volume} {107}},\ \bibinfo {pages} {097001} (\bibinfo {year}
  {2011})%
  \bibAnnoteFile{NoStop}{pavan}%
\bibitem{SuperconductingCuxBi2Se3}%
  \BibitemOpen
  \bibfield{author}{%
  \bibinfo {author} {\bibfnamefont{Y.~S.}\ \bibnamefont{Hor}} \emph{et~al.},\
  }%
  \bibfield{journal}{%
  \Doi{10.1103/PhysRevLett.104.057001}{\bibinfo {journal} {Phys. Rev. Lett.}}\
  }%
  \textbf{\bibinfo {volume} {104}},\ \bibinfo {pages} {057001} (\bibinfo {year}
  {2010})%
  \bibAnnoteFile{NoStop}{SuperconductingCuxBi2Se3}%
\bibitem{CuxBi2Se3ARPES}%
  \BibitemOpen
  \bibfield{author}{%
  \bibinfo {author} {\bibfnamefont{L.~A.}\ \bibnamefont{Wray}} \emph{et~al.},\
  }%
  \bibfield{journal}{%
  \Doi{10.1038/nphys1762}{\bibinfo {journal} {Nat Phys}}\ }%
  \textbf{\bibinfo {volume} {6}},\ \bibinfo {pages} {855} (\bibinfo {year}
  {2010}),\ ISSN \bibinfo {issn} {1745-2473}%
  \bibAnnoteFile{NoStop}{CuxBi2Se3ARPES}%
\bibitem{SCZhangBi2X3DiracCone}%
  \BibitemOpen
  \bibfield{author}{%
  \bibinfo {author} {\bibfnamefont{H.}~\bibnamefont{Zhang}} \emph{et~al.},\ }%
  \bibfield{journal}{%
  \Doi{10.1038/nphys1270}{\bibinfo {journal} {Nature Physics}}\ }%
  \textbf{\bibinfo {volume} {5}},\ \bibinfo {pages} {438} (\bibinfo {year}
  {2009})%
  \bibAnnoteFile{NoStop}{SCZhangBi2X3DiracCone}%
\bibitem{HgTeTI}%
  \BibitemOpen
  \bibfield{author}{%
  \bibinfo {author} {\bibfnamefont{C.}~\bibnamefont{Brune}}, \bibinfo {author}
  {\bibfnamefont{C.~X.}\ \bibnamefont{Liu}}, \bibinfo {author}
  {\bibfnamefont{E.~G.}\ \bibnamefont{Novik}}, \bibinfo {author}
  {\bibfnamefont{E.~M.}\ \bibnamefont{Hankiewicz}}, \bibinfo {author}
  {\bibfnamefont{H.}~\bibnamefont{Buhmann}}, \bibinfo {author}
  {\bibfnamefont{Y.~L.}\ \bibnamefont{Chen}}, \bibinfo {author}
  {\bibfnamefont{X.~L.}\ \bibnamefont{Qi}}, \bibinfo {author}
  {\bibfnamefont{Z.~X.}\ \bibnamefont{Shen}}, \bibinfo {author}
  {\bibfnamefont{S.~C.}\ \bibnamefont{Zhang}},\ and\ \bibinfo {author}
  {\bibfnamefont{L.~W.}\ \bibnamefont{Molenkamp}},\ }%
  \bibfield{journal}{%
  \bibinfo {journal} {Phys. Rev. Lett.}\ }%
  \textbf{\bibinfo {volume} {106}},\ \bibinfo {pages} {126803} (\bibinfo {year}
  {2011})%
  \bibAnnoteFile{NoStop}{HgTeTI}%
\bibitem{Heusler}%
  \BibitemOpen
  \bibfield{author}{%
  \bibinfo {author} {\bibfnamefont{S.}~\bibnamefont{Chadov}}, \bibinfo {author}
  {\bibfnamefont{X.}~\bibnamefont{Qi}}, \bibinfo {author}
  {\bibfnamefont{J.}~\bibnamefont{K��?bler}}, \bibinfo {author}
  {\bibfnamefont{G.~H.}\ \bibnamefont{Fecher}}, \bibinfo {author}
  {\bibfnamefont{C.}~\bibnamefont{Felser}},\ and\ \bibinfo {author}
  {\bibfnamefont{S.~C.}\ \bibnamefont{Zhang}},\ }%
  \bibfield{journal}{%
  \bibinfo {journal} {Nature Materials}\ }%
  \textbf{\bibinfo {volume} {9}},\ \bibinfo {pages} {541} (\bibinfo {year}
  {2010})%
  \bibAnnoteFile{NoStop}{Heusler}%
\bibitem{dai2008}%
  \BibitemOpen
  \bibfield{author}{%
  \bibinfo {author} {\bibfnamefont{X.}~\bibnamefont{Dai}}, \bibinfo {author}
  {\bibfnamefont{T.~L.}\ \bibnamefont{Hughes}}, \bibinfo {author}
  {\bibfnamefont{X.-L.}\ \bibnamefont{Qi}}, \bibinfo {author}
  {\bibfnamefont{Z.}~\bibnamefont{Fang}},\ and\ \bibinfo {author}
  {\bibfnamefont{S.-C.}\ \bibnamefont{Zhang}},\ }%
  \bibfield{journal}{%
  \bibinfo {journal} {Phys. Rev. B}\ }%
  \textbf{\bibinfo {volume} {77}},\ \bibinfo {pages} {125319} (\bibinfo {year}
  {2008})%
  \bibAnnoteFile{NoStop}{dai2008}%
\bibitem{FuKaneTIInversion}%
  \BibitemOpen
  \bibfield{author}{%
  \bibinfo {author} {\bibfnamefont{L.}~\bibnamefont{Fu}}\ and\ \bibinfo
  {author} {\bibfnamefont{C.~L.}\ \bibnamefont{Kane}},\ }%
  \bibfield{journal}{%
  \Doi{10.1103/PhysRevB.76.045302}{\bibinfo {journal} {Phys. Rev. B}}\ }%
  \textbf{\bibinfo {volume} {76}},\ \bibinfo {pages} {045302} (\bibinfo {year}
  {2007})%
  \bibAnnoteFile{NoStop}{FuKaneTIInversion}%
\bibitem{Novik:2005fk}%
  \BibitemOpen
  \bibfield{author}{%
  \bibinfo {author} {\bibfnamefont{E.~G.}\ \bibnamefont{Novik}}, \bibinfo
  {author} {\bibfnamefont{A.}~\bibnamefont{Pfeuffer-Jeschke}}, \bibinfo
  {author} {\bibfnamefont{T.}~\bibnamefont{Jungwirth}}, \bibinfo {author}
  {\bibfnamefont{V.}~\bibnamefont{Latussek}}, \bibinfo {author}
  {\bibfnamefont{C.~R.}\ \bibnamefont{Becker}}, \bibinfo {author}
  {\bibfnamefont{G.}~\bibnamefont{Landwehr}}, \bibinfo {author}
  {\bibfnamefont{H.}~\bibnamefont{Buhmann}},\ and\ \bibinfo {author}
  {\bibfnamefont{L.~W.}\ \bibnamefont{Molenkamp}},\ }%
  \bibfield{journal}{%
  \bibinfo {journal} {Physical Review B}\ }%
  \textbf{\bibinfo {volume} {72}},\ \bibinfo {pages} {035321} (\bibinfo {month}
  {07}\ \bibinfo {year} {2005})
  \bibAnnoteFile{NoStop}{Novik:2005fk}%
\bibitem{Chiu:2011uq}%
  \BibitemOpen
  \bibfield{author}{%
  \bibinfo {author} {\bibfnamefont{C.-K.}\ \bibnamefont{Chiu}}, \bibinfo
  {author} {\bibfnamefont{M.~J.}\ \bibnamefont{Gilbert}},\ and\ \bibinfo
  {author} {\bibfnamefont{T.~L.}\ \bibnamefont{Hughes}},\ }%
  \bibfield{journal}{%
  \bibinfo {journal} {Physical Review B}\ }%
  \textbf{\bibinfo {volume} {84}},\ \bibinfo {pages} {144507} (\bibinfo {month}
  {10}\ \bibinfo {year} {2011})
  \bibAnnoteFile{NoStop}{Chiu:2011uq}%
\bibitem{CdGMSCVortexStates64}%
  \BibitemOpen
  \bibfield{author}{%
  \bibinfo {author} {\bibfnamefont{C.}~\bibnamefont{Caroli}}, \bibinfo {author}
  {\bibfnamefont{P.~G.}\ \bibnamefont{{De~Gennes}}},\ and\ \bibinfo {author}
  {\bibfnamefont{J.}~\bibnamefont{Matricon}},\ }%
  \bibfield{journal}{%
  \Doi{10.1016/0031-9163(64)90375-0}{\bibinfo {journal} {Physics Letters}}\ }%
  \textbf{\bibinfo {volume} {9}},\ \bibinfo {pages} {307 } (\bibinfo {year}
  {1964}),\ ISSN \bibinfo {issn} {0031-9163}%
  \bibAnnoteFile{NoStop}{CdGMSCVortexStates64}%
\bibitem{KitaevWireMajorana01}%
  \BibitemOpen
  \bibfield{author}{%
  \bibinfo {author} {\bibfnamefont{A.~Y.}\ \bibnamefont{Kitaev}},\ }%
  \bibfield{journal}{%
  \Doi{10.1070/1063-7869/44/10S/S29}{\bibinfo {journal} {Sov. Phys.--Uspeki}}\
  }%
  \textbf{\bibinfo {volume} {44}},\ \bibinfo {pages} {131} (\bibinfo {year}
  {2001})%
  \bibAnnoteFile{NoStop}{KitaevWireMajorana01}%
\end{thebibliography}
\end{document}